\documentclass[12pt,preprint]{aastex}

\def\ltsima{$\; \buildrel < \over \sim \;$}

\def\lsim{\lower.5ex\hbox{\ltsima}}
\def\gtsima{$\; \buildrel > \over \sim \;$}
\def\gsim{\lower.5ex\hbox{\gtsima}}

\begin{document}
\title{Gravitational Lensing of the SDSS High-Redshift Quasars}

\author{J. Stuart B. Wyithe\altaffilmark{1} and Abraham Loeb}
\affil{Harvard-Smithsonian Center for Astrophysics, 60 Garden St.,
Cambridge, MA 02138;\\
swyithe@cfa.harvard.edu, aloeb@cfa.harvard.edu}

\altaffiltext{1}{Hubble Fellow}

\begin{abstract}

We predict the effects of gravitational lensing on the color-selected
flux-limited samples of $z_{\rm s}\sim4.3$ and $z_{\rm s}\ga5.8$ quasars, recently
published by the Sloan Digital Sky Survey (SDSS). Our main findings are:
{\bf (i)} The lensing probability should be 1--2 orders of magnitude higher
than for conventional surveys. The expected fraction of multiply-imaged
quasars is highly sensitive to redshift and the uncertain slope of the
bright end of the luminosity function, $\beta_h$. For $\beta_h=2.58$ (3.43)
we find that at $z_{\rm s}\sim 4.3$ and $i^*<20.0$ the fraction is $\sim 4\%$
($13\%$) while at $z_{\rm s}\sim6$ and $z^*<20.2$ the fraction is $\sim 7\%$
($30\%$).  {\bf (ii)} The distribution of magnifications is heavily skewed;
sources having the redshift and luminosity of the SDSS $z_{\rm s}\ga5.8$ quasars
acquire median magnifications of $med(\mu_{obs})\sim1.1$--$1.3$ and mean
magnifications of $\langle\mu_{obs}\rangle\sim5$--$50$. Estimates of the
quasar luminosity density at high redshift must therefore filter out
gravitationally-lensed sources.  {\bf (iii)} The flux in the Gunn-Peterson
trough of the highest redshift ($z_{\rm s}=6.28$) quasar is known to be
$f_{\lambda}<3\times10^{-19}{\rm erg\,sec^{-1}\,cm^{-2}\AA^{-1}}$. Should
this quasar be multiply imaged, we estimate a 40\% chance that light from
the lens galaxy would have contaminated the same part of the quasar
spectrum with a higher flux. Hence, spectroscopic studies of the epoch of
reionization need to account for the possibility that a lens galaxy, which
boosts the quasar flux, also contaminates the Gunn-Peterson trough.  {\bf
(iv)} Microlensing by stars should result in $\sim 1/3$ of multiply imaged
quasars in the $z_{\rm s}\ga5.8$ catalog varying by more than 0.5 magnitudes 
over the next decade. The median emission-line equivalent width of multiply
imaged quasars would be lowered by $\sim 20\%$ with respect to the intrinsic 
value due to differential magnification of the continuum and emission-line 
regions.

\end{abstract}


\section{Introduction}

The Sloan Digital Sky Survey (SDSS; Fukugita et al~1996; Gunn et al.~1998;
York et al.~2001) has substantially increased the number of quasars known
at a redshift $z_{\rm s}>3.5$ (Fan et al.~2000; Fan et al.~2001a,b,c; Schmidt et
al.~2001; Anderson et al.~2001). In this paper we study the effects of
gravitational lensing on two color-selected flux-limited samples of SDSS
quasars published by Fan et al.~(2000; 2001a,b,c). The first is a sample of
39 luminous quasars with redshifts in the range $3.6<z_{\rm s}<5.0$ (median of
$\sim4.3$) selected at a magnitude limit $i^*<20.0$. The second sample was
selected at a magnitude limit $z^*<20.2$ out of $i^*$-band dropouts
($i^*-z^*>2.2$) and consists of 4 quasars with redshifts $z_{\rm s}\ga5.8$ [one
of these, SDSS 1044-0125, was later found to have $z_{\rm s}=5.73$ (Djorgovski,
Castro, Stern \& Mahabal~2001)], including the most distant quasar known at
$z_{\rm s}=6.28$. These samples are very important for studies of quasar
evolution and early structure formation (e.g. Turner~1991; Haiman \&
Loeb~2001) as well as the ionizing radiation field at high redshift
(e.g. Madau, Haardt \& Rees~1999; Haiman \& Loeb~1998). In this paper we
examine the effects of gravitational lensing by foreground galaxies on the
observed properties of the SDSS quasars.

The significance of galaxy lensing for the statistics of very luminous quasars
was pioneered by Ostriker \& Vietri~(1986), while the importance of 
gravitational lensing for high redshift samples has been
emphasized by Barkana \& Loeb~(2000), who made specific predictions for
future observations by the Next Generation Space Telescope (planned for
launch in 2009\footnote{http://ngst.gsfc.nasa.gov/}). Among {\it existing}
samples, the SDSS high-redshift samples are unique in that they are likely
to yield a very high lensing probability.  This follows from two
trends. First, the lensing optical depth rises towards higher redshifts
(Turner~1991; Barkana \& Loeb~2000). More importantly, extrapolations of
the quasar luminosity evolution indicate that the SDSS limiting magnitude
is several magnitudes brighter than the luminosity function break. The fact
that the entire $z_{\rm s}\sim4.3$ and $z_{\rm s}\ga 5.8$ quasar 
samples reside in the part of the
luminosity function with a steep slope results in a very high magnification
bias (Turner~1980; Turner, Ostriker \& Gott~1984). This situation stands in
contrast to the typical survey at redshifts $z_{\rm s}\la 3$ for which the
limiting magnitude is fainter than the break magnitude at $m_B\sim19$. In
addition to having multiply-imaged sources, the high magnification bias in
SDSS should result in a high spatial correlation of high-redshift quasars
with foreground galaxies. Moreover, we expect some of these quasars to be
microlensed by the stellar populations of the lens galaxies; this should
result in variability of both the flux and emission-line equivalent widths
of the quasars (Canizares 1982).

The outline of the paper is as follows. In \S~\ref{Models} we summarize the
lens models and the assumed quasar luminosity function, and present the
formalism for calculating the expected distribution of magnifications due
to gravitational lensing. In \S~\ref{lensfrac} we calculate the fraction of
multiply-imaged sources and discuss their effect on the observed luminosity
function. In \S \ref{magobs} we discuss the distribution of magnifications
for the high-redshift quasar samples and in \S\ref{GPT} we find the level
of flux contamination within the Gunn-Peterson trough (Gunn \& Peterson~1965; 
Becker et al. 2001)
of a $z_{\rm s}\sim6$ lensed quasar due to the emission by its foreground lens
galaxy. Finally \S\ref{microlens} discusses the variability in flux and in
the distribution of equivalent widths due to microlensing. Throughout the
paper we assume a flat cosmology having density parameters of
$\Omega_m=0.35$ in matter, $\Omega_\Lambda=0.65$ in a cosmological
constant, and a Hubble constant $H_0=65{\rm \,km\,sec^{-1}\,Mpc^{-1}}$.

\section{Lens Models and Quasar Luminosity Function}
\label{Models}

\subsection{Lens Population}
We consider the probability for gravitational lensing by a constant
co-moving density of early type (elliptical/S0) galaxies which comprise
nearly all of the lensing optical depth (Kochanek~1996). The lensing rate
for an evolving (Press-Schechter~1974) population of lenses differs only by
$\la10\%$ (Barkana \& Loeb~2000) and is not considered. 
The distribution of velocity dispersions for the galaxy population is
described by a Schechter function with a co-moving density of early type
galaxies $n_\star = 0.27\times10^{-2}{\rm \,Mpc^{-3}}$ (Madgwick et
al.~2001), and the Faber-Jackson (1976) relation with an index of
$\gamma=4$. We adopt $\sigma_\star=220~{\rm km\,sec^{-1}}$ for the velocity
dispersion of an $L_\star$ galaxy, and assume that the dark matter velocity
dispersion equals that of the stars\footnote{A ratio ${\sigma_{\rm DM}\over
\sigma_{\rm stars}}=\sqrt{\frac{3}{2}}$ was introduced by Turner, Ostriker
\& Gott~(1984) as a correction factor for the simplest dynamical models
having a dark matter mass distribution with a radial power-law slope of -2
but a stellar distribution with a slope of -3. Kochanek~(1993, 1994) has
shown that $\sigma_{\rm DM}=\sigma_{\rm stars}$ instead results in image
separations consistent with those observed, and isothermal mass
profiles that produce dynamics consistent with local early type galaxies.}  
$\sigma_{\rm DM}=\sigma_{\rm stars}$. 
Because the high redshift quasars are color selected, foreground 
galaxies that would be detected by SDSS
must be removed from the population of potential lens
galaxies. In this paper we assume that lens galaxies having $i^*_{gal}>22.2$
will result in a high redshift quasar missing the color selection cuts.
This value is 2.2 magnitudes fainter than the $i^*$ limit of the $z_{\rm s}\sim4.3$ 
survey, and 2 magnitudes fainter than the $z^*$ limit of the $z_{\rm s}\ga5.8$
survey (corresponding to the $i^*$-band dropout condition $i^*-z^*>2.2$).
We calculate the apparent magnitude of a galaxy with
velocity dispersion $\sigma$ at redshift $z_{\rm d}$ to be
\begin{equation}
\label{eqn1}
i^* \approx M_{i^*} -10 \log_{10}\left(\frac{\sigma}{\sigma_\star}\right) +
K(z_{\rm d}) +2.5 \log_{10}\left[10^{-0.6z_{\rm d}}\right]
+ 5 \log_{10}\left(\frac{d_L(z_{\rm d})}{10}\right),
\end{equation}
where $M_{i^*}$ is the $i^*$-band absolute magnitude of an $L_\star$ early
type galaxy (Madgwick et al.~2001; and color corrections from Fukugita,
Shimasaku \& Ichikawa~1995 and Blanton et al.~2001), $d_L$ is the
luminosity distance in parsecs and $K(z_{\rm d})$ is the $k$-correction (from Fukugita, 
Shimasaku \& Ichikawa~1995). The fourth term in equation~(\ref{eqn1}) is 
derived from the evolution of mass to rest frame B-band luminosity 
ratios of lens galaxies [$\frac{d\log(M/L)}{dz_{\rm d}}=-0.6$; 
Koopmans \& Treu~(2001)]. The main results of this paper are only weakly dependent
on the detailed surface brightness and color evolution\footnote{Equation~(\ref{eqn1}) 
assumes no color evolution of early type galaxies. For galaxies at $z_{\rm d}\sim0.8$, $i^*$-band 
corresponds approximately to rest-frame $B$-band, and equation~(\ref{eqn1}) is insensitive 
to the assumption of no color evolution. At lower galaxy redshifts $i^*$-band corresponds to wavelengths longer than $B$-band in the rest frame. Since the stellar population 
becomes redder as it ages, the assumption of no color evolution underestimates the 
apparent magnitude of galaxies at $z_{\rm d}\la0.8$, resulting in conservative lens statistics.} 
of the lens galaxies.

\begin{figure*}[hptb]
\epsscale{.7}
\plotone{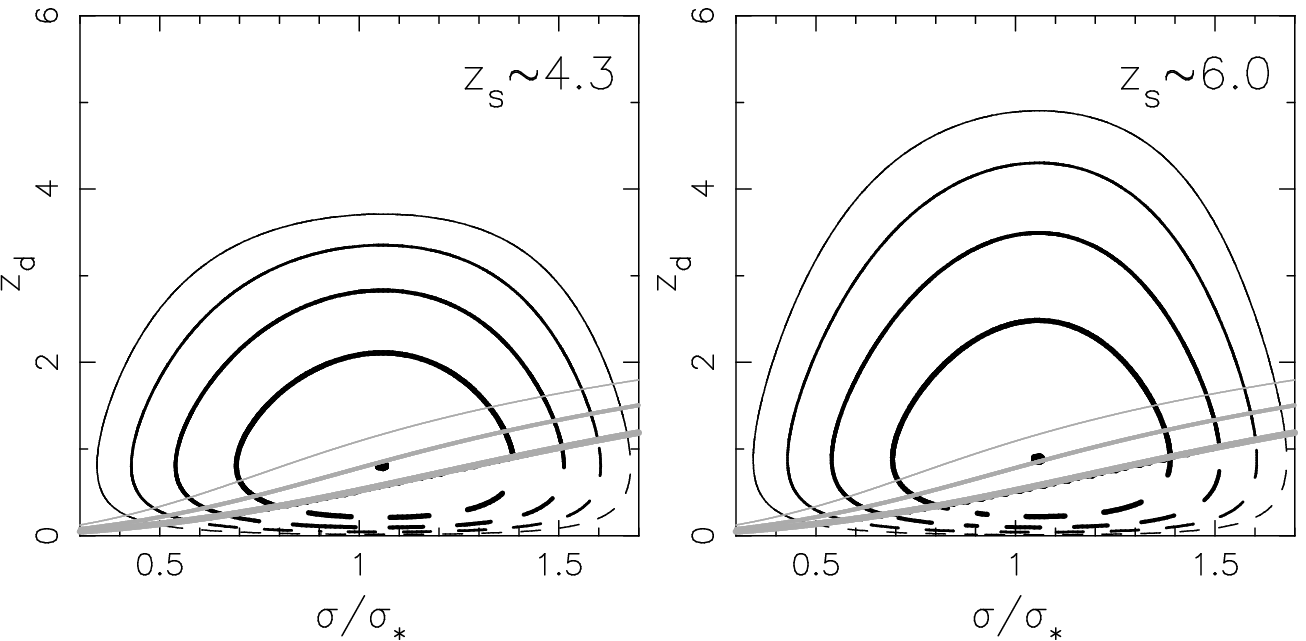}
\caption{\label{fig1}Contours of joint probability of multiple image optical depth
for $\sigma/\sigma_\star$ and $z_{\rm d}$ assuming lens galaxies fainter than 
$i^*_{gal}=22.2$ (solid contours), and for all lens galaxies 
(dashed contours). The dots are placed at the distribution mode, and the contours 
are plotted at 1/3, 1/9, 1/27 and 1/81 the peak height. The thick grey line shows 
the locus of galaxies having $i^*_{gal}=22.2$, the thinner grey line 
$i^*_{gal}=23.2$ and the thinnest grey line $i^*_{gal}=24.2$.}
\end{figure*}

\begin{figure*}[hptb]
\epsscale{.7}
\plotone{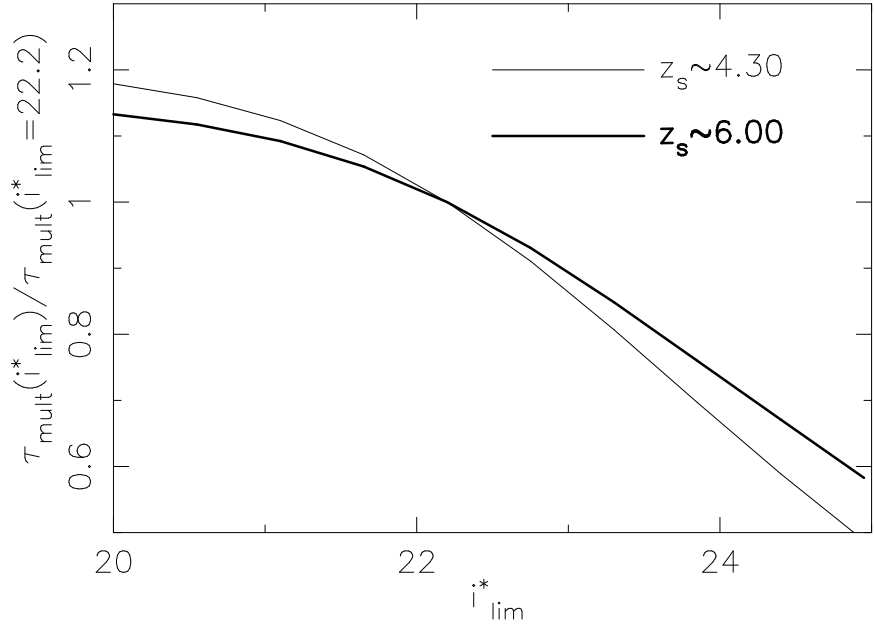}
\caption{\label{fig2}The multiple image optical depth $\tau_{mult}$ for different 
bright lens galaxy limits $i^*_{gal}$, normalized by the optical depth 
for a limit of $i^*_{gal}=22.2$.}
\end{figure*}

Figure~\ref{fig1} shows the joint probability contours of multiple image optical depth
for $\sigma/\sigma_\star$ and $z_{\rm d}$ assuming sources at
redshifts of $z_{\rm s}\sim4.3$ (left) and $z_{\rm s}\sim6.0$ (right). The solid
contours represent the undetected lens galaxies which are included in
calculations of the lensing statistics through the remainder of this paper,
and the dashed lines represent the rest of the potential lens galaxy 
population. The $i^*_{gal}=22.2$
borderline that separates these populations of galaxies is also shown. 
The fraction of the lens population lost due to the requirement of
non-detectability of the lens galaxy is 20--30\%. Figure~\ref{fig1} also shows
loci corresponding to $i^*_{gal}=23.2$ and 24.2. These give an indication
of the effect on lens statistics of disregarding still fainter lens galaxies.
As a further guide to the sensitivity
of the lens statistics on the assumed bright lens galaxy limit, we have computed 
the optical depth for different bright lens galaxy limits and plotted them as
a fraction of the optical depth for a limit of  $i^*_{gal}=22.2$ in Figure~\ref{fig2}. 
This figure may be used to estimate the variation of the lensing probabilities
calculated in this paper with the absence of lens galaxies brighter than
different limiting values $i^*_{gal}$. For example, excluding lens galaxies 
down to $i^*_{gal}=23.2$ will reduce the multiple imaging rates presented in
this paper to $\sim80-90\%$ of their quoted values. In addition to the
aforementioned samples we will also consider for reference the statistics
of lensing by the entire E/S0 galaxy population of quasars with $z_{\rm s}=2.1$.

The galaxy mass distribution is modeled as a
combination of stars and a smooth dark matter halo having in total a mass
profile of a singular isothermal sphere (SIS). The stars are distributed
according to de-Vaucouleurs profiles having characteristic radii and 
surface brightnesses determined as a function of $\sigma_{star}$ from the 
relations of Djorgovski \& Davis~(1987). A constant mass-to-light ratio
as a function of radius is assumed. The inclusion of stars allows us to calculate
three possible effects due to microlensing: (i) a possible increase in the
magnification of bright quasars; (ii) short-term variability of the quasar
flux; and (iii) changes in the equivalent width of the broad emission lines
in the quasar spectrum due to differential magnification of the continuum
and emission-line regions. We compute the microlensing statistics by
convolving a large number of numerical magnification maps with the
distribution of microlensing optical depth and shear for lines of sight to
random source positions (Wyithe \& Turner~2002). We assume a source size of
$10^{15}{\rm cm}$ (corresponding to 10 Schwarzschild radii of a
$3\times10^8M_{\odot}$ black hole) and microlens masses of $0.1M_{\odot}$.
The microlens surface mass density is evolved with redshift in proportion
to the fraction of the stellar mass that had formed by that redshift.  We
assume a constant star formation rate at $z>1$ and a rate proportional to
$(1+z)^{3}$ at $z<1$ (Hogg 1999 and references therein; Nagamine, Cen \&
Ostriker~2000). The mass-to-light ratios were normalized so that the
elliptical/S0 plus spiral populations at redshift zero contain a cosmological
density parameter in stars of $\Omega_\star=0.005$ (Fukugita, Hogan \&
Peebles~1998). The model for elliptical galaxies therefore has two
characteristic angular scales. On arcsecond scales the mass distribution
follows the radial profile of an isothermal sphere (which determines the
macrolensing cross-section). On micro-arcsecond scales, the grainy mass
distribution of the stars yields different phenomena related to
microlensing.

The inclusion of non-sphericity in the lenses is beyond the scope of this
paper.  Although previous studies (Kochanek \& Blandford~1987; Blandford \&
Kochanek~1987) found that the introduction of ellipticities $\la 0.2$ into
nearly singular profiles has little effect on the lensing cross-section and
image magnification, the strong magnification bias will favor a high
fraction of 4-image lenses (Rusin \& Tegmark~2001) as well as an increase
in the number of multiple image systems. One consequence of this effect
will be to double the expected microlensing rate (see
\S\ref{microlens}). We also do not include source extinction by the lens
galaxy, which should arise primarily in the rarer spiral galaxy lenses.
Spiral galaxies may be more common at the higher lens redshifts encountered
for the high redshift quasar catalogs, and should be considered in future
extensions of this work.

\subsection{Magnification Distribution}
\label{magdist}

\begin{figure*}[hptb]
\epsscale{1.}  
\plotone{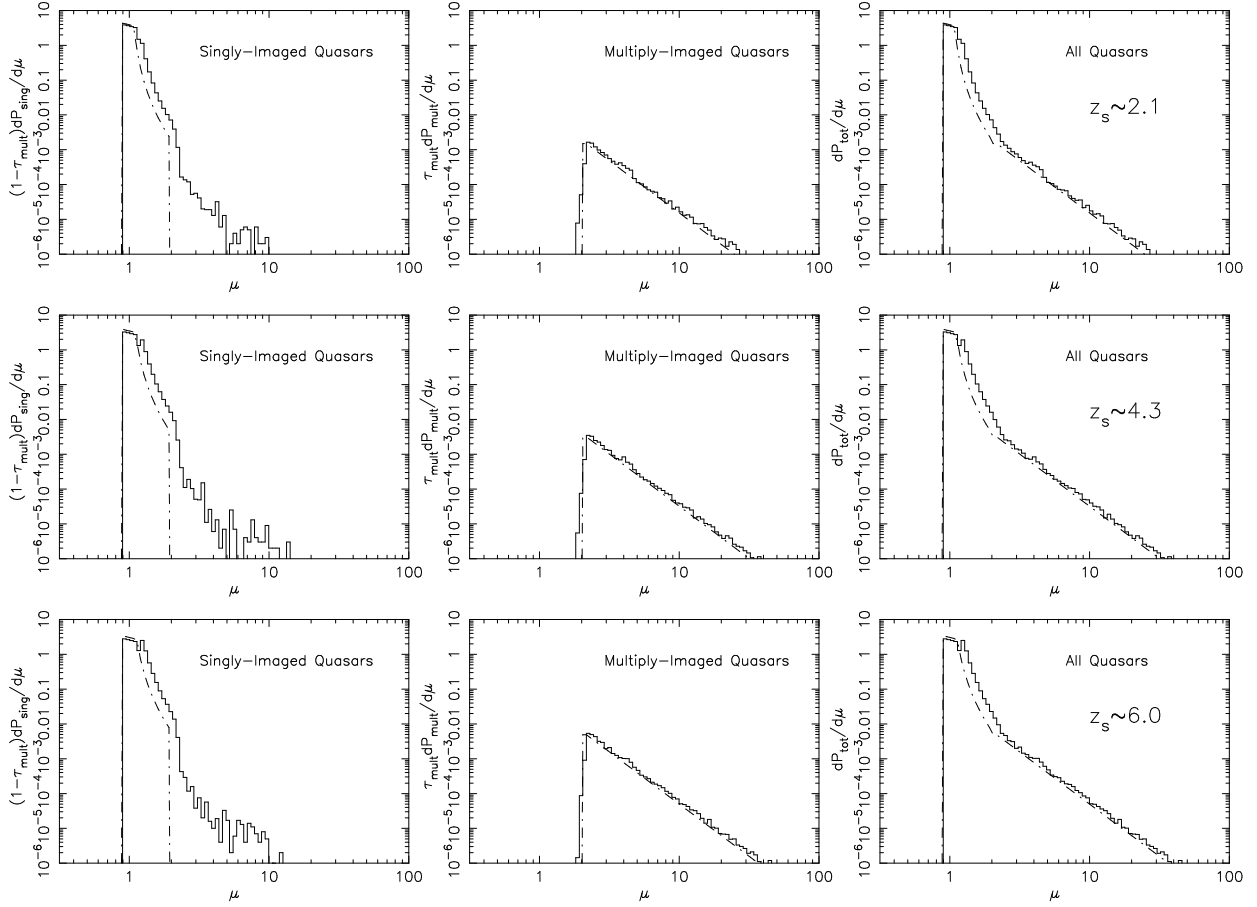}
\caption{\label{fig3} Magnification distributions for quasars. The left,
center and right panels show distributions for singly-imaged quasars, for
the sum of images of multiply-imaged quasars, and for the combination of
the two (i.e. all quasars). The top, central and lower rows show cases
where $z_{\rm s}\sim2.1$, $z_{\rm s}\sim4.3$ and $z_{\rm s}\sim6.0$.}
\end{figure*}

We define $dP/d\mu$ as the normalized differential probability per unit magnification
and $\tau_{mult}$ as the multiple-image optical depth.  The magnification
distributions were computed for singly-imaged quasars
[$(1-\tau_{mult})\frac{dP_{sing}}{d\mu}$], multiply-imaged quasars
[$\tau_{mult}\frac{dP_{mult}}{d\mu}$, where $\mu$ is the sum of the
multiple image magnifications], and all quasars [$\frac{dP_{tot}}{d\mu}=
(1-\tau_{mult}) \frac{dP_{sing}}{d\mu}+\tau_{mult}\frac{dP_{mult}}{d\mu}$].
These are illustrated in Figure~\ref{fig3} for $z_{\rm s}=2.1$ (top
column), $z_{\rm s}=4.3$ (central column) and $z_{\rm s}=6.0$ (lower column). The
histograms show the spread resulting from microlensing around the usual SIS
distributions (which are shown for comparison by the dot-dashed lines),
including a non-zero probability for single images with magnifications
larger than 2, and a nonzero probability for the total magnification of a
multiply-imaged source to be smaller than 2. The method described in Wyithe
\& Turner~(2002) does not determine the magnification along all lines of
sight. Specifically, those source positions whose lines of sight have
microlensing optical depth $\kappa<10^{-4}$ and hence magnifications near 1
are not considered. However the average magnification for a random source
position must be unity, and we add probability smoothly to the single image
distribution in the bins between 0.9 and 1.1 such that 
[$\frac{dP_{tot}}{d\mu}$] has unit mean and [$(1-\tau_{mult})\frac{dP_{sing}}{d\mu}$] 
is normalized to $1-\tau_{mult}$. The detail of the treatment of the 
distributions near $\mu=1$ has no barring on the resulting lens statistics. 

In order to find the fraction of multiply-imaged sources, the magnification
distributions need to be convolved with the quasar luminosity function. We
discuss the luminosity function next.

\subsection{Quasar Luminosity Function}
\label{lumfunc}

The standard double power-law luminosity function (Boyle, Shanks \&
Peterson~1988; Pei~1995) for the number of quasars per comoving volume per
unit luminosity
\begin{equation}
\label{LF}
\phi(L,z_{\rm s})=\frac{\phi_\star/L_\star(z_{\rm s})}{[L/L_\star(z_{\rm s})]^{\beta_l}+[L/L_\star(z_{\rm s})]^{\beta_h}}
\end{equation}
provides a successful representation of the observed quasar luminosity
function at redshifts $\la 3$. The functional dependence on redshift is in
the break luminosity $L_\star$ indicating pure luminosity evolution. At
$z_{\rm s}\ll 3$ the break luminosity evolves as a power-law in redshift and the
number counts increase with redshift; higher redshift surveys indicate that
there is a decline in the space density of bright quasars beyond $z_{\rm s}\sim3$
(Warren, Hewett \& Osmer~1994; Schmidt, Schneider \& Gunn~1995; Kennefick,
Djorgovski \& de Carvalho~1995). Madau, Haardt \& Rees~(1999) suggested an
analytic form for the evolution of $L_\star$ of the form
\begin{equation}
\label{Lstar}
L_\star(z_{\rm s})=L_{\star,0}(1+z_{\rm s})^{-(1+\alpha)}\frac{e^{\zeta z_{\rm s}}(1+e^{\xi z_\star})}{e^{\xi
z_{\rm s}}+e^{\xi z_\star}},
\end{equation} 
where $\alpha$ is the slope of the power-law continuum of quasars (taken to
be -0.5 throughout this paper). Fan et al.~(2001b) found from their sample
of SDSS quasars at $z_{\rm s}\sim4.3$ that the slope of the bright end of the
luminosity function has evolved to $\beta_h\sim2.58$ from the $z_{\rm s}\sim3$
value of $\beta_h\sim3.43$. This result supported the findings of Schmidt,
Schneider \& Gunn~(1995). We therefore consider two luminosity functions in
this study. In one case we consider $\beta_h=3.43$ below $z_{\rm s}=3$ and 
$\beta_h\sim2.58$ above $z_{\rm s}=3$. As a second case, we assume that the 
slope of the bright end $\beta_h=3.43$ does not evolve above $z_{\rm s}\sim3$ 
and so we adopt pure luminosity evolution at all redshifts.
We emphasize that while $\beta_h=3.43$ is found not to vary at $z_{\rm s}\la3$,
the (small) samples of quasars at $z_{\rm s}\ga4$ suggest the flatter slope 
($\beta_h=2.58$) for the bright end of the luminosity function. Thus 
currently a luminosity function having $\beta_h=2.58$ best fits the
high redshift data. In the remainder of the paper, discussions of numerical
results will list those for $\beta_h=2.58$ first.

With the above two prescriptions, the evolution of $L_\star$ is adjusted to
adequately describe the low redshift luminosity function (Hartwick \&
Schade~1990), and the space density of quasars at $z_{\rm s}\sim4.3$ and
$z_{\rm s}\sim6$ measured by Fan et al.~(2001b,c). The resulting luminosity
functions at $z_{\rm s}\sim2.1$, $z_{\rm s}\sim4.3$ and $z_{\rm s}\sim6.0$ are plotted in the
upper panels of Figure~\ref{fig4} and the corresponding parameters for
equations~(\ref{LF}) and (\ref{Lstar}) are listed in Table~\ref{tab1}. The
dotted line shows the unlensed luminosity function and the solid line shows
the lensed luminosity function (e.g. Pei~1995b) including
microlensing. Note that for $\beta_h=3.43$ the number density of quasars at
$z_{\rm s}\sim6.0$ with $M_{B}<-27.6$ is increased by a factor of $\sim1.5$ due to
lensing. For the computation of lens statistics for flux limited samples of 
quasars we use the cumulative
version of equation~(\ref{LF}), namely $N(>L,z_{\rm s}) =\int_L^\infty
dL'\phi(L',z_{\rm s})$.

\section{Multiple Imaging Rates}
\label{lensfrac}

We now combine the magnification distributions described in
\S\ref{magdist} with the luminosity functions of \S\ref{lumfunc} to find
the fraction of multiply imaged quasars,
\begin{equation}
\label{eqn4}
F_{\rm MI}(z_{\rm s})=\frac{\int_{0}^{\infty}d\mu'\tau_{\rm mult}\frac{dP_{\rm
mult}}{d\mu'}N(>\frac{L_{\rm
lim}}{\mu'},z_{\rm s})}{\int_{0}^{\infty}d\mu'\left[\tau_{\rm mult}\frac{dP_{\rm
mult}}{d\mu'}+ (1-\tau_{\rm mult})\frac{dP_{\rm
sing}}{d\mu'}\right]N(>\frac{L_{\rm lim}}{\mu'},z_{\rm s})},
\end{equation} 
where the multiple image optical-depth 
obtains values of $\tau_{\rm mult}= 0.0019$, 0.0040 and 0.0059 for
$z_{\rm s}=2.1$, 4.3 and 6.0 respectively, and $L_{lim}$ is the limiting
luminosity at a redshift $z_{\rm s}$ corresponding to the limiting survey
magnitude\footnote{We use a limiting magnitude for simplicity and assume
that all multiply-imaged systems will be identified in high-resolution
follow-up observations. However a full calculation to interpret observed
statistics should include selection functions for both the inclusion of the
quasar in the survey (Fan et al.~2001a,c) with and without additional light 
from a foreground lens galaxy, and for the detection of
multiple images in follow-up observations (Turner, Ostriker \& Gott~1984).}
$m_{lim}$.  The limiting luminosity $L_{\rm lim}$ was determined from
$m_{lim}$ using the luminosity distance for the assumed cosmology and a
$k$-correction computed from a model quasar spectrum including the mean absorption
by the intergalactic medium (M$\o$ller \& Jakobsen~1990; Fan~1999). Note 
that equation~(\ref{eqn4}) reduces to the usual approximation for magnification bias
in the limit of small $\tau_{mult}$ and shallow luminosity functions.

\begin{figure*}[hptb]
\epsscale{.8}
\plotone{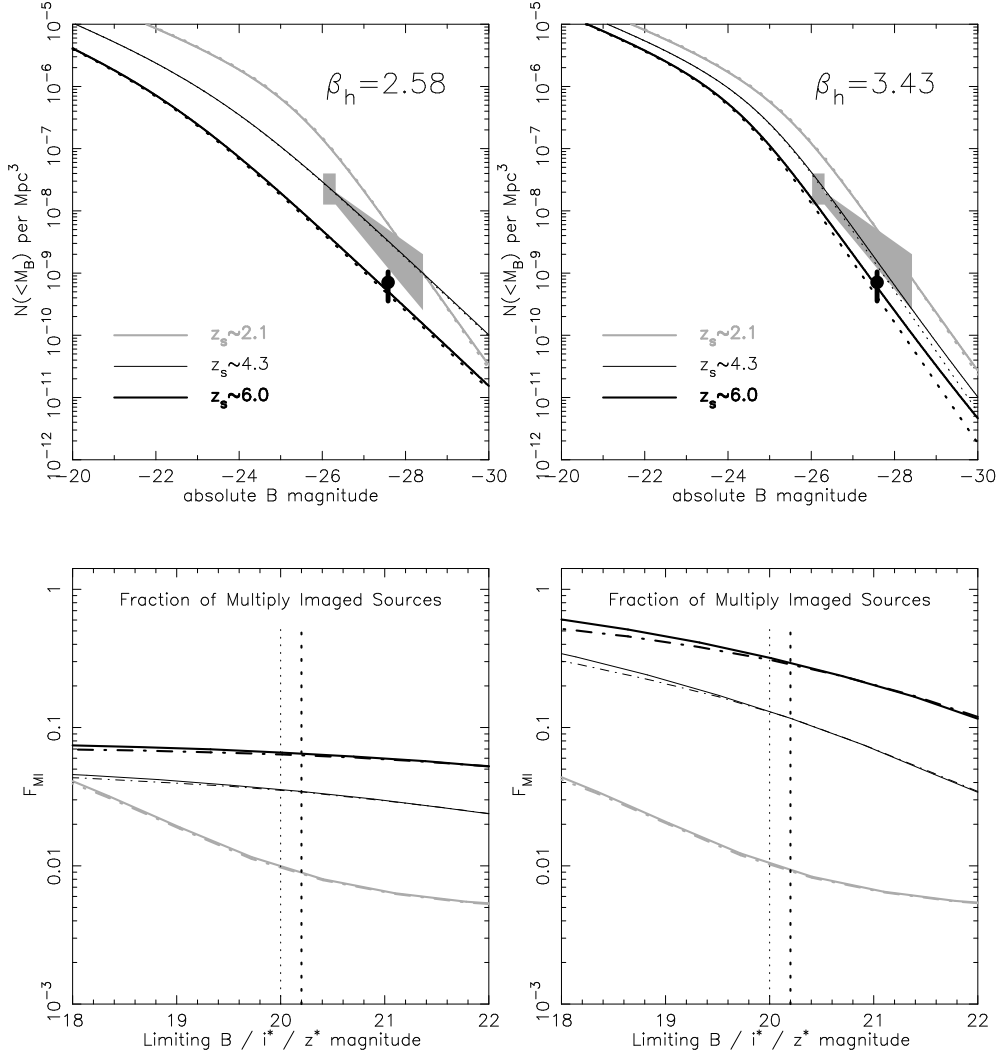}
\caption{\label{fig4} {\it Upper panels:} The cumulative quasar luminosity
functions at $z_{\rm s}\sim2.1$ (thick light lines), $z_{\rm s}\sim4.3$ (thin dark
lines) and $z_{\rm s}\sim6.0$ (thick dark lines). The solid and dotted lines show
the lensed and intrinsic luminosity functions, respectively. The solid grey
region represents the non-parametric luminosity function from the SDSS
$z_{\rm s}\sim4.3$ quasar sample (Fan et al.~2001b), and the dark point with
vertical error bar shows the measured space density of quasars at
$z_{\rm s}\sim6.0$ (Fan et al.~2001c). The left and right panels show luminosity
functions with $\beta_h=3.43$ for $z_{\rm s}<3$ but $\beta_h=2.58$ and
$\beta_h=3.43$ respectively for $z_{\rm s}>3$. {\it Lower panels:} The fraction,
$F_{\rm MI}$, of multiply-imaged quasars with redshifts $z_{\rm s}\sim2.1$, $4.3$
and $6.0$, as a function of limiting $B$, $i^*$ and $z^*$ magnitudes,
respectively. The solid and dot-dashed lines correspond to calculations
that include microlensing and those that only consider a smooth lensing
mass distribution. The limiting magnitudes are marked by the vertical
dotted lines, and the line types are the same as in the upper panels.}
\end{figure*}

The lower panels of Figure~\ref{fig4} show the predicted lens fraction for
samples of quasars at $z_{\rm s}\sim2.1$ (thick light lines), $z_{\rm s}\sim4.3$ (thin
dark lines) and $z_{\rm s}\sim6.0$ (thick dark lines) as a function of the
limiting survey magnitude. The solid lines show the fraction found when
microlensing is included while the dot-dashed lines show results for pure
SIS lenses. Results have been computed for the two luminosity functions
discussed in \S 2.3. The vertical lines mark the limiting magnitudes for
the high redshift samples, namely $i^*\sim20.0$ for the $z_{\rm s}\sim4.3$ survey
and $z^*\sim20.2$ for the $z_{\rm s}\ga5.8$ survey. The magnification bias and
hence the multiple image fraction is significantly higher for brighter
limiting magnitudes. The lens fraction asymptotes to a constant value at
the brighter limits considered here since these are sufficiently
brighter than the break magnitude that only very large and hence rare
magnifications could result in the inclusion of quasars fainter than
$L_\star$. Therefore at these bright limits the luminosity function can be
considered as a single power-law with a resulting bias that is insensitive
to the limiting magnitude. Obviously a steeper bright end slope leads to a
larger magnification bias and a higher multiple image fraction. We find
that a typical survey of quasars at low redshift ($z_{\rm s}\sim2.1$) has 
$F_{\rm MI}\sim0.01$ (for a limiting B-magnitude of $m_B=20$).  This 
estimate is consistent with the measured lens fraction in the 
HST Snapshot Survey (Maoz et al.~1993).  However as
argued by Barkana \& Loeb~(2000), bright high redshift quasar surveys
should find much higher values of $F_{\rm MI}$. For $\beta_h=2.58$ we find
$F_{\rm MI}\sim0.04$ at $z_{\rm s}\sim4.3$ and $F_{\rm MI}\sim0.07$ at
$z_{\rm s}\sim6.0$, respectively. For $\beta_h=3.43$ the fractions are even
higher, $F_{\rm MI}\sim0.13$ at $z_{\rm s}\sim4.3$ and $F_{\rm MI}\sim0.30$ at
$z_{\rm s}\sim6.0$ respectively. Note the inferred absolute quasar luminosity is 
quite sensitive to the $k$-correction, particularly
from the absorption spectrum since quasars at $z_{\rm s}\sim6$ are nearly
completely absorbed blueward of Ly$\alpha$. As a result the inferred lens
fraction for a fixed apparent magnitude limit is also sensitive to the 
$k$-correction. However we find that microlensing makes
little difference to the multiple imaging fraction unless the survey limit
is very bright.

\begin{figure*}[hptb]
\epsscale{.8}
\plotone{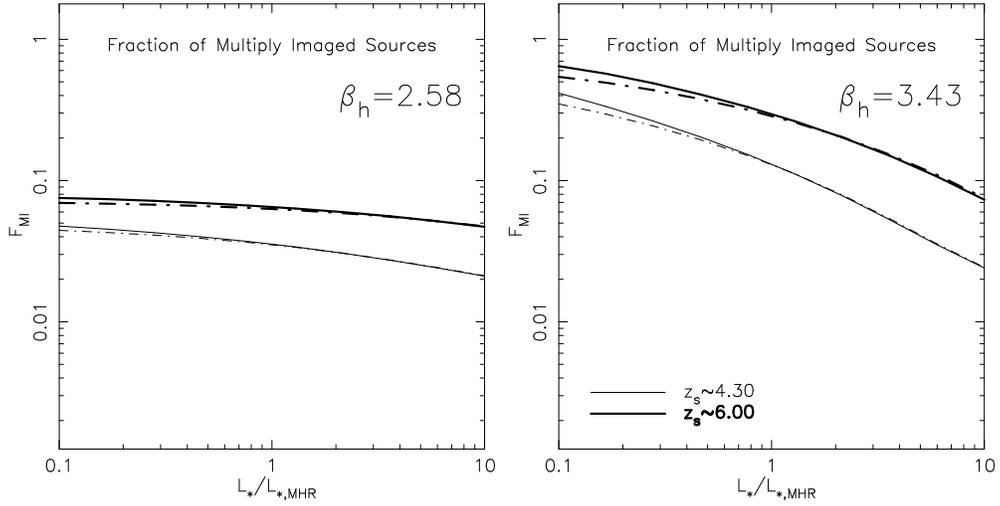}
\caption{\label{fig5}The fraction, $F_{\rm MI}$, of multiply-imaged quasars
$z_{\rm s}\sim4.3$ (thin dark lines) and $z_{\rm s}\sim6.0$ (thick dark lines) for 
limiting magnitudes of $i^*=20.0$ and $z^*=20.2$ respectively, as a 
function of the break
luminosity in units of the value quoted in Table~\ref{tab1}. The left and
right panels show luminosity functions with $\beta_h=3.43$ for $z_{\rm s}<3$ but
$\beta_h=2.58$ and $\beta_h=3.43$ respectively for $z_{\rm s}>3$. The solid and
dot-dashed lines correspond to calculations that include microlensing and
those that only consider a smooth lensing mass distribution,
respectively. }
\end{figure*}

We have extrapolated the luminosity function at $z_{\rm s}\sim4.3$ and $z_{\rm s}\sim6$
from the well-studied luminosity function at lower redshifts. We now
consider how the results may be affected should this extrapolation be
invalid. Conversely, given the large multiple image fraction it may be
possible in the future to use the observed fraction of multiple images in
these samples to constrain properties of the luminosity function (modulo
systematics in lens modeling and cosmology). 
As a demonstration of this potential use, we
calculate multiple image fractions for sources at $z_{\rm s}\sim4.3$ with $i^*<20.0$
and at $z_{\rm s}\sim6.0$ with $z^*<20.2$ assuming values of break luminosity that
are within factors of 10 above and below the corresponding values of
$L_\star$ listed in Table~\ref{tab1}. The results are plotted in
Figure~\ref{fig5} (line types as in Figure~\ref{fig4}).  We see that the
multiple image fraction is rather insensitive to the location of the
break for the shallow ($\beta_h=2.58$) luminosity function. However the
location of the break plays a more significant role for the steep
($\beta_h=3.43$) luminosity function, particularly at $z_{\rm s}\sim4.3$ where
the limiting magnitude is closer to the extrapolated location of the break.

\section{Magnification of Observed Sources}
\label{magobs}

In \S\ref{magdist} we considered the probability of magnification along
lines of sight to random sources.  In this section we expand on the discussion 
in Wyithe \& Loeb~(2002) and  calculate the
\textit{a-posteriori} probability for the magnification of a known quasar
in the high-redshift samples. If a quasar is observed with a magnification
of $\mu_{obs}$, then it is intrinsically fainter by a factor of
$\mu_{obs}$ and therefore more abundant by a factor of
$[\phi(L/\mu_{obs})/(\mu_{obs}\phi(L))]$. The distribution of magnifications observed
at redshift $z_{\rm s}$ in a flux-limited sample is
\begin{equation}
\label{obsmageqn1}
\frac{dP}{d\mu_{\rm obs}} = \frac{\left[\tau_{\rm mult}\frac{dP_{\rm
mult}}{d\mu}\vert_{\mu=\mu_{\rm obs}} + (1-\tau_{\rm mult})\frac{dP_{\rm
sing}}{d\mu}\vert_{\mu=\mu_{\rm obs}}\right]N(>\frac{L_{\rm lim}}{\mu_{\rm
obs}})}{\int_0^{\infty}d\mu'\left[\tau_{\rm mult} \frac{dP_{\rm
mult}}{d\mu}\vert_{\mu=\mu'} + (1-\tau_{\rm mult})\frac{dP_{\rm
sing}}{d\mu}\vert_{\mu=\mu'}\right]N(>\frac{L_{\rm lim}}{\mu'})}.
\end{equation}
The bias in this equation introduces additional skewness into the
magnification distribution. This effect is particularly severe for the
high-redshift samples in which quasars are selected to be bright (so that
they reside in the steep part of the luminosity function).

\begin{figure*}[hptb]
\epsscale{.8}
\plotone{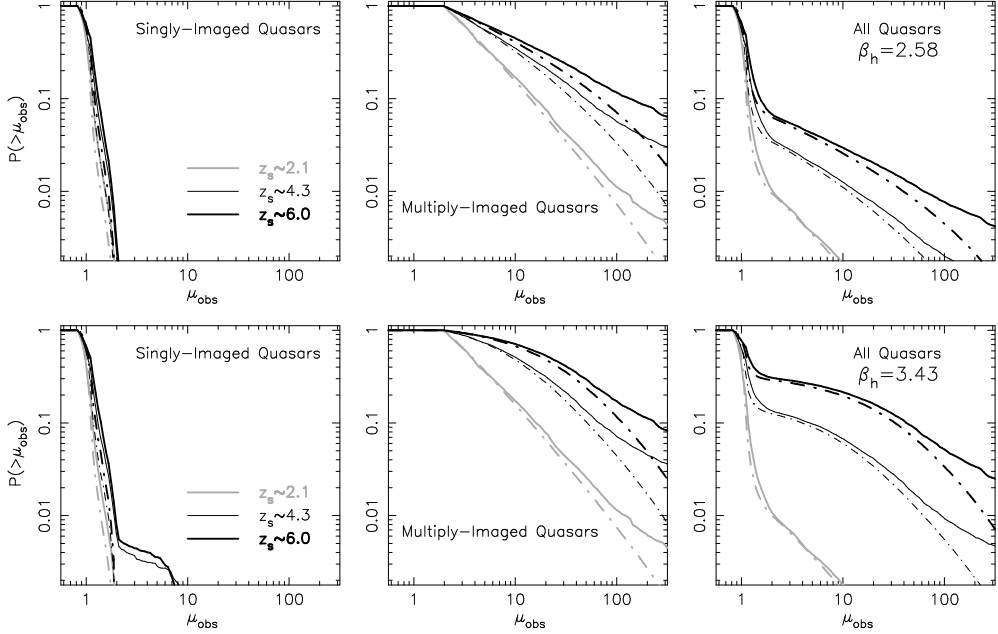}
\caption{\label{fig6} The probability for observing a magnification larger
than $\mu_{\rm obs}$ of quasars in flux-limited samples. The left, center
and right panels show distributions for singly-imaged quasars, for the sum
of images of a multiply-imaged quasar, and for the combination of the two
(i.e. all quasars). The upper and lower panels show results for luminosity
functions with $\beta_h=3.43$ for $z_{\rm s}<3$ but $\beta_h=2.58$ and
$\beta_h=3.43$ respectively for $z_{\rm s}>3$. In each panel we show the
distributions at quasar redshifts of $z_{\rm s}\sim2.1$ (thick light lines),
$z_{\rm s}\sim4.3$ (thin dark lines) and $z_{\rm s}\sim6.0$ (thick dark lines) for
limiting magnitudes of $m_B=20.0$, $i^*=20.0$ and $z^*=20.2$
respectively. The solid and dot-dashed lines correspond to calculations
that include microlensing and those that only consider a smooth lensing
mass distribution, respectively.  }
\end{figure*}

Figure~\ref{fig6} shows the cumulative distributions of observed
magnifications for sources at redshifts $z_{\rm s}\sim2.1$, $4.3$ and $6.0$ (line
types as in Figure~\ref{fig4}) in samples of quasars with $m_B<20.0$,
$i^*<20.0$ and $z^*<20.2$ respectively. Distributions are shown for single
images, for the sum of multiple images and for all images.  The upper 
and lower panels show
results for luminosity functions with $\beta_h=2.58$ and $\beta_h=3.43$ (at
$z_{\rm s}>3$), respectively. The distributions are highly skewed, having medians
near unity but means as high as a few tens (values are listed in
Table~\ref{tab2}). Note that with $\beta_h=3.43$, multiple images generate
a fairly flat distribution out to high magnifications for the $z_{\rm s}\sim4.3$
and $z_{\rm s}\sim6.0$ samples.  This follows from the fact that at luminosities
above $L_\star$ the integrated luminosity function is nearly as steep as
the high magnification tail of the magnification distribution.  As a result
of microlensing the single image distributions show a small probability for
magnifications larger than 2. Microlensing also causes a significant
increase in the probability of observing the very largest magnifications in
multiply-imaged sources. Highly magnified multiple images are more likely
in the high redshift samples because of the larger magnification bias. This
trend combined with the fact that multiple images are more common in
general at high redshift results in a very large difference between the
probabilities of observing $\mu_{\rm obs}>10$ at the different redshifts
considered.

\begin{figure*}[hptb]
\epsscale{.8}
\plotone{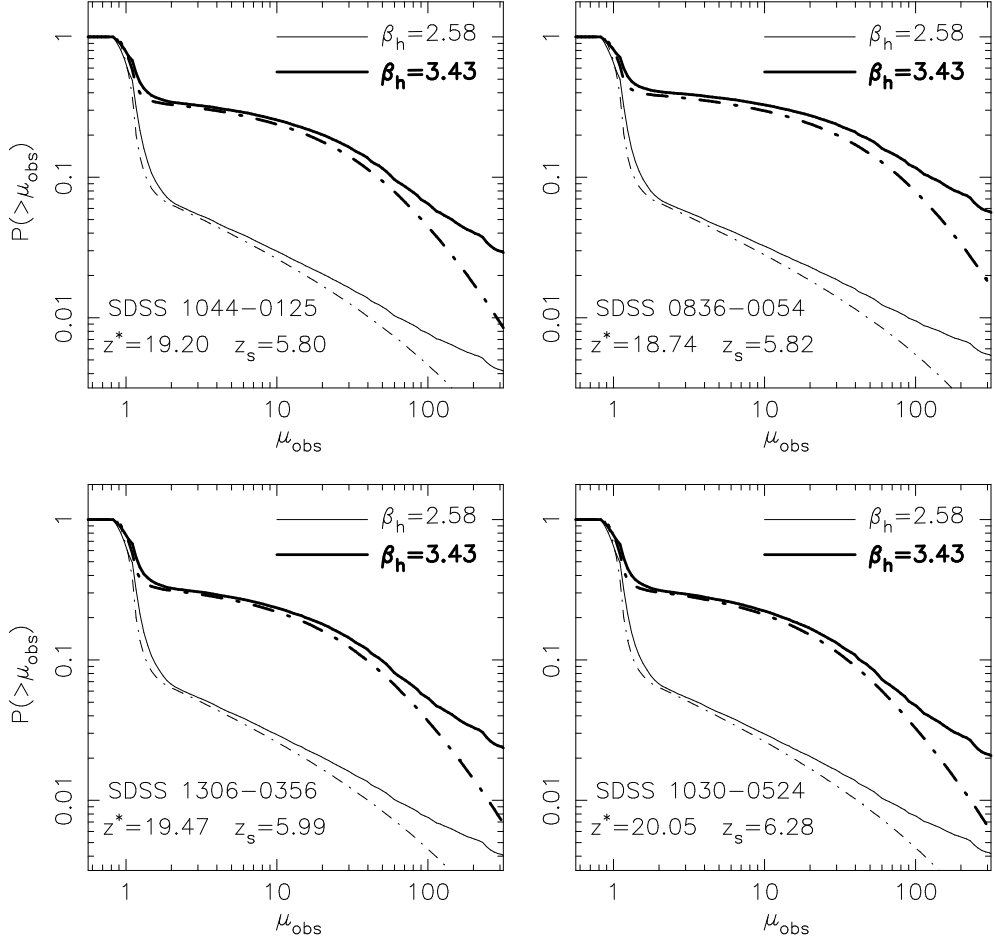}
\caption{\label{fig7} The probability of observing a magnification larger
than $\mu_{\rm obs}$ for quasars having the fluxes and redshifts of the
four SDSS $z_{\rm s}\ga5.8$ quasars. The thin and thick lines show results for
luminosity functions with $\beta_h=3.43$ for $z_{\rm s}<3$ but $\beta_h=2.58$ and
$\beta_h=3.43$ respectively for $z_{\rm s}>3$. The solid and dot-dashed lines
correspond to calculations that include microlensing and those that only
consider a smooth lensing mass distribution, respectively. In
Table~\ref{tab3} we list the multiple image probability, $F_{\rm MI}$, and
the distribution mean, $\langle\mu_{obs}\rangle$, and median,
$med(\mu_{obs})$, for each of these quasars.  }
\end{figure*}

The \textit{a-posteriori} multiple image fraction and distribution of
magnifications observed for sources having a luminosity $L$ and a redshift
$z_{\rm s}$ are
\begin{equation}
F_{\rm MI}(z_{\rm s})=\frac{\int_{0}^{\infty}d\mu'\tau_{\rm mult}\frac{dP_{\rm
mult}}{d\mu'}\frac{\phi(L/\mu',z_{\rm s})}{\mu'}}{\int_{0}^{\infty}d\mu'
\left[\tau_{\rm mult}\frac{dP_{\rm mult}}{d\mu'}+ (1-\tau_{\rm
mult})\frac{dP_{\rm sing}}{d\mu'}\right]\frac{\phi(L/\mu',z_{\rm s})}{\mu'}},
\end{equation} 
and
\begin{equation}
\frac{dP}{d\mu_{\rm obs}} = \frac{\left[\tau_{\rm mult}\frac{dP_{\rm
mult}}{d\mu}\vert_{\mu=\mu_{\rm obs}} + (1-\tau_{\rm mult})\frac{dP_{\rm
sing}}{d\mu}\vert_{\mu=\mu_{\rm obs}}\right]\frac{\phi(L/\mu_{\rm
obs})}{\mu_{obs}}}{\int_0^{\infty}d\mu'\left[\tau_{\rm mult} \frac{dP_{\rm
mult}}{d\mu}\vert_{\mu=\mu'} + (1-\tau_{\rm mult})\frac{dP_{\rm
sing}}{d\mu}\vert_{\mu=\mu'}\right]\frac{\phi(L/\mu')}{\mu'}}.
\end{equation}
We have calculated these quantities for the 4 SDSS $z_{\rm s}\ga5.8$
quasars discovered by Fan et al.~(2000; 2001c). The results are plotted in
Figure~\ref{fig7}, and the resulting medians, means and multiple imaging
fractions are listed in Table~\ref{tab3}. Assuming $\beta_h=2.58$ (3.43)
the probability for multiple imaging is $F_{\rm MI}\sim0.06-0.07$
($0.3-0.4$) for these quasars, with expected magnifications of
$\langle\mu_{\rm obs}\rangle\sim4.5-5.5$ ($23-50$). Thus, if the bright
end of the luminosity function is shallow ($\beta_h=2.58$) as suggested by
Fan et al.~(2001b) then we do not expect any lenses among the existing
$z_{\rm s}\ga5.8$ sample. On the other hand, if $\beta_h=3.43$ (as
extrapolated from the pure luminosity evolution observed at low redshifts)
then we expect one or two out of the four $z_{\rm s}\ga5.8$ quasars to be
multiply imaged and magnified by a large factor, while the others should
have magnifications $\la 2$.

\section{Lens Galaxy Light and the Gunn-Peterson Trough}
\label{GPT}

The spectra of quasars at $z_{\rm s}\sim6$ provide an exciting probe of the epoch
of reionization. The spectrum of the very highest redshift quasar
discovered to date was found by Becker et al.~(2001) to have higher than
expected neutral hydrogen absorption, indicating a possible Gunn-Peterson
trough due to the pre-ionized intergalactic medium (Gunn \& Peterson~1965).
While we have required that lens galaxies not be detected by the
SDSS imaging survey, it is possible that lens galaxies would contribute
flux in deeper follow-up observations. The high fraction of lensed quasars
expected in the SDSS catalog at $z_{\rm s}\sim 6$ implies that light from the
lens galaxy may contaminate the Gunn-Peterson trough for a substantial
fraction of all quasars. This has the potential to limit the ability of
deep spectroscopic observations to probe the evolution of the neutral
hydrogen fraction during the epoch of reionization in some cases.

\begin{figure*}[hptb]
\epsscale{.5}
\plotone{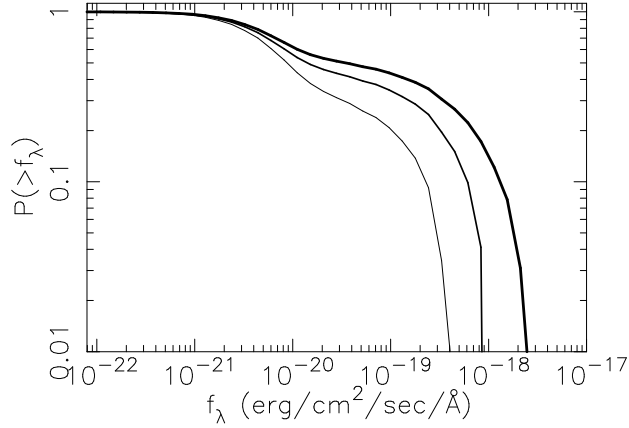}
\caption{\label{fig8}The probability that a lens galaxy responsible for a
multiply imaged quasar will contribute a flux greater than $f_\lambda$ in
the Gunn-Peterson trough of a quasar at $z_{\rm s}\sim6$. Three lines are shown 
corresponding to lens galaxies fainter than $i^*_{gal}=22.2$, 23.2 and 24.2
(thicker lines denote brighter magnitude limits).}
\end{figure*}

We have estimated the distribution of flux per unit wavelength,
$f_\lambda$, in the Gunn-Peterson trough due to lens galaxy light for
multiply-imaged $z_{\rm s}\sim 6$ quasars as a function of lens galaxy velocity
dispersion and redshift, and convolved the result with the joint
probability distribution shown in Figure~\ref{fig1}. We used the $i^*$-band
absolute magnitude of an $L_\star$ early-type galaxy (Madgwick et al.~2001;
and color corrections from Fukugita, Shimasaku \& Ichikawa~1995 and Blanton
et al.~2001) and computed the apparent flux assuming the
Faber-Jackson~(1976) relation, the mean galaxy spectrum from
Kennicutt~(1992), and a rest-frame 
mass-to-light ratio that evolves as $\frac{dlog(M/L)}{dz}=-0.6$ 
(Koopmans \& Treu~2001). We further assumed that all of the galaxy light is added
to the quasar spectrum. Figure~\ref{fig8} shows the probability that flux
at a level greater than $f_{\lambda}$ will be observed in the Gunn-Peterson
trough. Probabilities are shown assuming lens galaxies fainter than
$i^*_{gal}=22.2$, 23.2 and 24.2.
The spectra of the $z_{\rm s}=6.28$ quasar published by Becker et
al.~(2001) and Pentericci~(2001) show that the flux in the Gunn-Peterson
trough is $\la 3\times10^{-19}{\rm \,erg\,sec\,cm^{-2}\,\AA}$. We estimate
that $\sim40\%$ of lens galaxies should contribute flux above this level 
($i^*_{gal}=22.2$). Hence,
the observed flux limit in the Gunn-Peterson trough does not rule out
lensing for this object. On the other hand the probability is not
negligible and demonstrates the need to account for the possibility of a
contaminating lens galaxy.

\section{Microlensing and the High-redshift Quasar Samples}
\label{microlens}

So far, we have found that the rare high magnifications caused by
microlensing due to stars do not significantly affect the multiple image
fraction, but do result in a significant excess of the highest
magnifications. We now consider more direct manifestations of
microlensing. In particular, we compute the fraction of all quasars that
should vary by more than $\Delta m$ magnitudes during a 10 year period, and
the expected microlensing-induced distortion of the distribution of
equivalent widths for the broad emission lines of quasars. The results in
this section have been computed assuming a luminosity function having
$\beta_h=3.43$ at all redshifts, which yield larger lensing fractions as 
discussed in previous sections. However we point out how the results can be 
applied to the case of $\beta_h=2.58$ where appropriate.

\subsection{Quasar Flux Variability due to Microlensing}

Microlensing induces quasar variability due to the relative angular motion
of the observer, lens and source. The variability occurs on a shorter
timescale for lens galaxies at lower redshifts due to the larger projected
transverse velocity. Since the SDSS high-redshift samples select against
low-redshift lens galaxies, we expect measurable microlensing variability
to be rare. On the other hand, the high magnification bias for these
samples suggests a high multiple imaging rate and a large galaxy-quasar
angular correlation, bringing more lines of sight to regions on the sky
with a significant microlensing optical depth. We have computed the
fraction of quasar images that vary by more than $\Delta m$ magnitudes
during the 10 years following discovery using the methods described in
Wyithe \& Turner~(2002). Magnification bias is calculated based on the
magnification of the first light-curve point (the sum of magnifications for
a multiply-imaged source). The effective transverse source plane velocity
was computed from the two transverse velocity components for each of the
source, lens and observer (Kayser, Refsdal \& Stabell~1986).  We assume
each component to be Gaussian-distributed with a dispersion of 400${\rm
\,km\,sec^{-1}}$.

\begin{figure*}[hptb]
\epsscale{.8}
\plotone{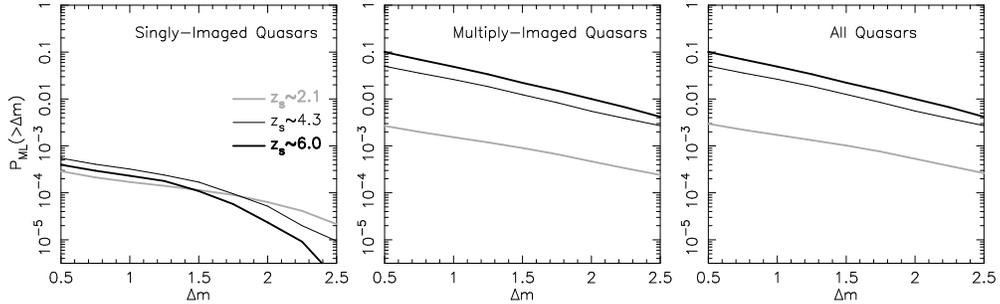}
\caption{\label{fig9}The fraction of quasars that are microlensed by more
than $\Delta m$ during the 10 years following their discovery. The left,
center and right panels show results for singly-imaged quasars, for the
images of a multiply-imaged quasar (with each image considered separately),
and for all quasars. The luminosity function is assumed to have
$\beta_h=3.43$ at all redshifts. In each panel we show results at quasar
redshifts of $z_{\rm s}\sim2.1$ (thick light lines), $z_{\rm s}\sim4.3$ (thin dark
lines) and $z_{\rm s}\sim6.0$ (thick dark lines) for limiting magnitudes of
$m_B=20.0$, $i^*=20.0$ and $z^*=20.2$ respectively.}
\end{figure*}

Figure~\ref{fig9} shows the resulting probabilities for the
variability amplitude of quasars at $z_{\rm s}\sim2.1$, $4.3$ and $6.0$, with
line types as in Figure~\ref{fig4}.  The fraction of quasars that are
singly-imaged and microlensed is shown in the left panel, the fraction of
quasar images (counting each image separately) that appear in multiple
image systems and are microlensed is shown in the central panel, and the
fraction of all quasar images to be microlensed is shown in the right
panel. Microlensing variability is dominated by multiple-image
systems. This is particularly true for the high redshift samples, where the
high magnification bias results in a large multiple-image fraction. While
only 1 in 300 images at $z_{\rm s}\sim2.1$ vary by more than $\Delta m=0.5$
magnitudes, we find that at $z_{\rm s}\sim6$ the fraction has risen to
$\sim10\%$. The fraction of quasars at $z_{\rm s}\sim6$ that are 
multiply-imaged and microlensed assuming the flat luminosity function 
having $\beta_h=2.58$ is approximately obtained by multiplying the above 
result by $0.07/0.30$ to correct for the multiple image rate.

Since the multiple image fraction is $\sim0.3$ we find that
$\sim1$ in 3 multiply-imaged quasars at $z_{\rm s}\sim6$ will show microlensing
of more than $\Delta m=0.5$ magnitudes in one of their images during the 10
years following discovery. Similarly we find that at $z_{\rm s}\sim4.3$,
$\sim5\%$ of quasars will vary by more than $\Delta m=0.5$
magnitudes. Since the multiple image fraction is $\sim0.13$, this indicates
that $\sim1$ in 3 multiply-imaged sources will exhibit microlensing above
this level in one of their images. These fractions hold if $\beta_h=2.58$. 
Quasars also vary
intrinsically. However, while microlensing causes independent variability
of the quasar images, intrinsic variability is observed in all images
separated by the lens time delay (e.g.~Kundic et al.~1997).  After the lens
time delay is determined, it should be possible to separate microlensing
variability from intrinsic variability.

The SDSS is expected to discover $\sim30$ quasars with $z_{\rm s}\ga5.8$ when
completed. If $\beta_h=3.43$, monitoring of these quasars should yield 
$\sim3$ microlensing events. Unlike lensed quasars with lens galaxies 
having $z_{\rm d}\ll1$, the
source to lens angular diameter distance ratio in these cases is near
unity. The source size is therefore comparable to the projected microlens
Einstein radius and so proposed caustic crossing experiments to map the
structure of the accretion disc (e.g. Agol \& Krolik~1999) will not be
applicable.  Nonetheless, the microlens Einstein radius still represents an
interesting characteristic scale, and the detection of microlensing would
provide an upper limit on the extent of the source size. If the source
emission is interpreted as originating from a smooth accretion disk, then
microlensing variability would place an upper limit on the mass of the
quasar black hole. This, in turn, would yield a lower limit on the ratio
between the quasar luminosity and its maximum (Eddington) value which would
be useful in constraining models for early structure formation (Haiman \&
Loeb 2001).

\subsection{Distortion in the Equivalent Width Distribution 
of Quasar Emission Lines}

\begin{figure*}[hptb]
\epsscale{.8}
\plotone{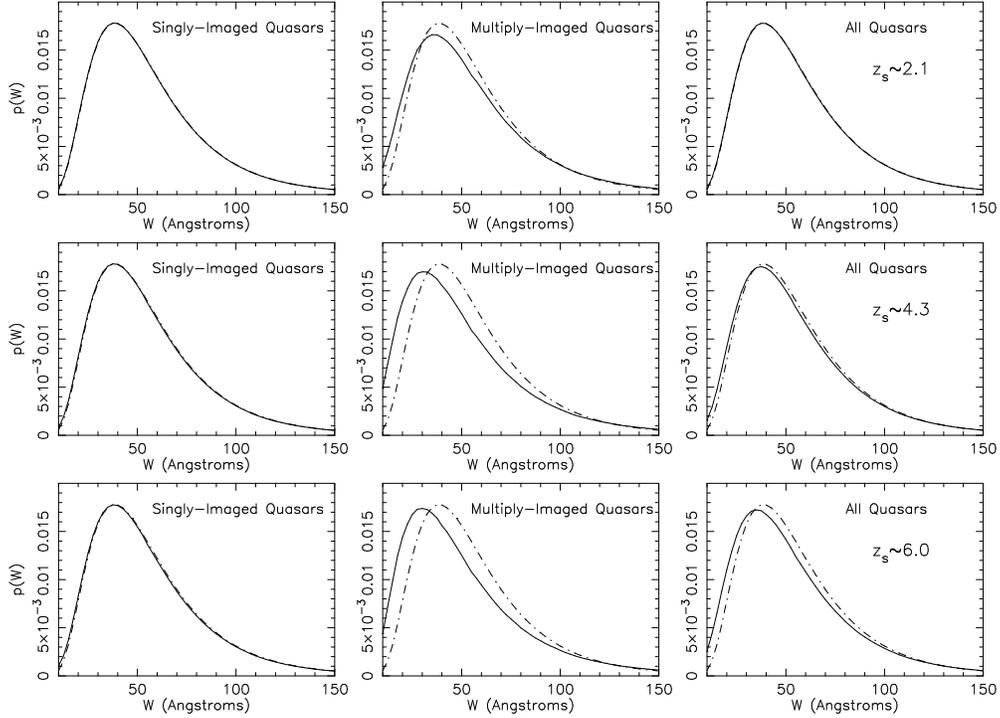}
\caption{\label{fig10}The microlensed distribution (solid lines) of
equivalent-widths ($W$) for the broad emission lines of quasars. The left, center
and right panels show results for singly-imaged quasars, for the sum of
images in multiply-imaged quasars, and for all quasars, respectively. The
luminosity function is assumed to have $\beta_h=3.43$ at all redshifts. The
upper, central and lower panels show results at quasar redshifts of
$z_{\rm s}\sim2.1$, $4.3$ and $6.0$ with limiting magnitudes of $m_B=20.0$,
$i^*=20.0$ and $z^*=20.2$, respectively. The dot-dashed lines show the
assumed intrinsic distribution.}
\end{figure*}

Fan et al.~(2001c) noted that their high redshift quasar ($z>5.8$) 
selection criteria favor objects with strong emission lines (particularly 
Ly$\alpha$). Since
microlensing results in variability of emission-line equivalent-widths by
differentially magnifying the compact continuum region compared to the more
extended region that produces the lines (Canizares 1982; Delcanton et
al.~1994; Perna \& Loeb~1998), it is important to quantify how large this
effect might be in the high-redshift quasar samples. We follow Perna \&
Loeb~(1998) and compute the distribution of relative magnifications
${dP}/{d(\mu/\langle\mu\rangle)}$ (where $\langle\mu\rangle$ is the mean
magnification near the line of sight) which is then convolved with the
intrinsic log-normal equivalent width distribution. We consider
magnification bias when constructing ${dP}/{d(\mu/\langle\mu\rangle)}$.
Microlensing is more likely to lower the observed equivalent width because
sources are more likely to be selected when their continuum magnification
is above the average. Results are shown in Figure~\ref{fig10} for the
$z_{\rm s}\sim2.1$ (upper row), $z_{\rm s}\sim4.3$ (central row), and
$z_{\rm s}\sim6.0$ (lower row) samples. The distributions for singly imaged
sources show almost no departure from the intrinsic distribution, while the
median of the multiply-imaged source distribution is lowered by $\sim10\%$,
$\sim20\%$ and $\sim30\%$ respectively for the $z_{\rm s}\sim2.1$, $4.3$, and
$6.0$ samples. These values are slightly reduced if $\beta_h=2.58$. Microlensing 
results in a net reduction in the median equivalent width 
for $z_{\rm s}\sim6.0$ quasars of $\sim15\%$ relative to its intrinsic value. Thus, we
expect the observed equivalent width distribution to be altered from its
intrinsic state, but the level of variation to be smaller than the intrinsic
spread. Hence, the distortion of equivalent widths due to microlensing
should not bias against the selection of lensed quasars in the sample of 
Fan et al.~(2000c).

\section{Summary}

We have shown that gravitational lensing should be more common by 1--2
orders of magnitude in the high-redshift quasar catalogs at $z_{\rm s}\sim4.3$
and $z_{\rm s}\ga5.8$ recently published by the Sloan Digital Sky Survey (SDSS),
as compared to previous quasar samples. The reasons for this large increase
are twofold. First, the optical depth for multiple imaging increases with
redshift. Second, all quasars in the SDSS high redshift catalogs populate
the bright end of the luminosity function making the magnification bias
stronger. Using extrapolations of the luminosity function having bright-end 
slopes of $-2.58$ (suggested by observations at $z_{\rm s}\sim4.3$) and 
$-3.43$ (found at all $z_{\rm s}\la3$) we find that multiple image
fractions of $\sim4\%$ $(13\%)$ for quasars at $z_{\rm s}\sim4.3$ (brighter than
$i^*=20.0$) and $\sim7\%$ ($30\%$) for quasars at $z_{\rm s}\sim6.0$ (brighter than
$z^*=20.2$). These fractions depend sensitively on the value of the break
luminosity and the bright-end slope. Thus, the observed lensing rate in
these bright samples can be used to constrain the parameters of the quasar
luminosity function at high redshifts.

We have computed the distribution of magnifications for quasars in flux
limited samples. A steep bright-end slope results in high probabilities for
very large magnifications because the slope of the luminosity function is
comparable to the slope of the high magnification tail of the magnification
distribution. Assuming bright end luminosity function slopes of $-2.58$ 
and $-3.43$, the median and mean magnifications are $med(\mu_{obs})=1.02$ (1.09)
and $\langle\mu_{obs}\rangle=2.09$ (6.00) for quasars at $z_{\rm s}\sim4.3$ 
(brighter than $i^*=20.0$), and $med(\mu_{obs})=1.09$ (1.19) and 
$\langle\mu_{obs}\rangle=4.55$ (23.0) for quasars at $z_{\rm s}\sim6.0$ (brighter 
than $z^*=20.2$).  The considerable abundance of systems with high 
magnifications implies that estimates of the quasar luminosity density need 
to be done with care after taking out gravitationally lensed systems.

Observations of the highest redshift quasar ($z_{\rm s}=6.28$) show a complete
Gunn-Peterson trough at a level $<3\times10^{-19}{\rm
\,erg\,sec\,cm^{-2}\,\AA}$. We find that $\sim40\%$ of multiple image lens
galaxies (fainter than $i^*_{gal}=22.2$) will contribute flux in the 
Gunn-Peterson trough (of a $z_{\rm s}\sim6$ quasar) at a level above 
$3\times10^{-19}{\rm \,erg\,sec\,cm^{-2}\,\AA}$. For some 
quasars the contamination of the Gunn-Peterson
trough with flux from lens galaxies may therefore limit the ability of deep
spectroscopic observations to probe the evolution of the neutral hydrogen
fraction during the epoch of reionization.

We have also computed microlensing statistics for high redshift quasars in
flux limited samples, and found that microlensing will be dominated by
multiply-imaged sources. One third of multiply-imaged quasars at
$z_{\rm s}\sim4.3$ (brighter than $i^*=20.0$) and one third of multiply-imaged
quasars at $z_{\rm s}\sim6.0$ (brighter than $z^*=20.2$) will vary due to
microlensing by more than 0.5 magnitudes during the decade following
discovery. This variability allows for the exciting possibility of using
differential microlensing magnification to probe the smallest scales of
bright quasars beyond a redshift of 6. Finally, we find that microlensing
lowers the median of the distribution of emission-line equivalent-widths
for multiply-imaged quasars at $z_{\rm s}\sim6.0$ (brighter than $z^*=20.2$) by
$\sim15\%$ relative to its intrinsic value. This effect is smaller than the
intrinsic spread and should therefore not bear on the quasar selection
function.

\acknowledgements 

The authors would like to thank Ed Turner, Chris Kochanek, Dave Rusin and 
Josh Winn for helpful discussions.
This work was supported in part by NASA grants NAG 5-7039, 5-7768, and NSF
grants AST-9900877, AST-0071019 for AL. JSBW is supported by a Hubble
Fellowship grant from the Space Telescope Science Institute, which is
operated by the Association of Universities for Research in Astronomy,
Inc., under NASA contract NAS 5-26555.

\newpage

\newpage

\begin{table}[htbp]
\caption{{\small Parameters for the luminosity function described by
equations~(\ref{LF}) and (\ref{Lstar}).}}
\vspace{3mm}
\begin{center}
\begin{tabular}{cccccccc}
\hline
$\beta_h$ $(z_{\rm s}>3)$   &  $\beta_h$ $(z_{\rm s}<3)$   &  $\beta_l$  &  $\phi_{\star}$ $({\rm Gpc^{-3}})$   &  $L_{\star,0}$ ($L_\odot$)   &  $z_{\star}$  & $\zeta$  &  $\xi$  \\ \hline
2.58       &3.43       &   1.64      & 624 &$1.50\times10^{11}$& 1.60          & 2.65     & 3.30    \\ 
3.43       &3.43       &   1.64      & 624 &$1.50\times10^{11}$& 1.45          & 2.70     & 2.90    \\\hline
\end{tabular}
\end{center}
\label{tab1}

\vspace{10mm}
\caption{{\small Multiple image fractions ($F_{\rm MI}$), mean
magnifications ($\langle\mu_{obs}\rangle$) and median magnifications
[$med(\mu_{obs})$] for samples of quasars with different source redshifts
($z_{\rm s}$) and limiting magnitudes ($m_{lim}$).}}
\vspace{3mm}
\begin{center}
\begin{tabular}{cccccccccc}
\hline & && \multicolumn{3}{c}{$\beta_h(z_{\rm s}>3)=2.58$} &&
          \multicolumn{3}{c}{$\beta_h(z_{\rm s}>3)=3.43$} \\ $z_{\rm s}$ & $m_{lim}$ &&
          $F_{\rm MI}$ & $\langle\mu_{obs}\rangle$ & $med(\mu_{obs})$ && $F_{\rm
          MI}$ & $\langle\mu_{obs}\rangle$ & $med(\mu_{obs})$ \\\hline 2.1 &
          $m_B=20.0$ & & 0.009 & 1.09 & 0.98 && 0.009 & 1.09 & 0.98 \\ 4.3 &
          $i^*=20.0$ && 0.036 & 2.09 & 1.02 && 0.13 & 6.00 & 1.09 \\ 6.0 &
          $z^*=20.2$ & & 0.065 & 4.55 & 1.09 && 0.30 & 23.0 & 1.19 \\\hline
\end{tabular}
\end{center}
\label{tab2}

\vspace{10mm}

\caption{{\small Multiple image fractions ($F_{\rm MI}$), mean
magnifications ($\langle\mu_{obs}\rangle$) and median magnifications
[$med(\mu_{obs})$] for the four SDSS $z_{\rm s}\ga5.8$ quasars with their
corresponding $z^*$ magnitudes.}}
\vspace{3mm}
\begin{center}
\begin{tabular}{ccccccccccc}
\hline & & && \multicolumn{3}{c}{$\beta_h(z_{\rm s}>3)=2.58$} &&
           \multicolumn{3}{c}{$\beta_h(z_{\rm s}>3)=3.43$} \\ & $z_{\rm s}$ & $z^*$ &&
           $F_{MI}$ & $\langle\mu_{obs}\rangle$ & $med(\mu_{obs})$ && $F_{\rm MI}$
           & $\langle\mu_{obs}\rangle$ & $med(\mu_{obs})$ \\\hline SDSS 1044-0125 &
           5.80 & 19.20 & & 0.07 & 4.76 & 1.09 && 0.34 & 29.8 & 1.22 \\
           SDSS 0836-0054 & 5.82 & 18.74 && 0.07 & 5.56 & 1.10 && 0.40 &
           49.8 & 1.28 \\ SDSS 1306-0356 & 5.99 & 19.47 & & 0.06 & 4.68 &
           1.09 && 0.32 & 25.4 & 1.20 \\\ SDSS 1030-0524 & 6.28 & 20.05 & &
           0.07 & 4.76 & 1.09 && 0.31 & 23.0 & 1.20 \\\hline
\end{tabular}
\end{center}
\label{tab3}
\end{table}

\end{document}